# ET PROBES: LOOKING HERE AS WELL AS THERE

**JOHN GERTZ***
*Zorro Productions, 125 University Avenue, Suite 101, Berkeley, CA 94710, USA.*
Email: jgertz@firsst.org

Almost all SETI searches to date have explicitly targeted stars in the hope of detecting artificial radio or optical transmissions. It is argued that extra-terrestrials (ET) might regard sending physical probes to our own Solar System as a more efficient means for sending large amounts of information to Earth. Probes are more efficient in terms of energy and time expenditures; may solve for the vexing problem of Drake's L factor term, namely, that the civilization wishing to send information may not co-exist temporally with the intended recipient; and they alleviate ET's reasonable fear that the intended recipient might prove hostile. It is argued that probes may be numerous and easier to find than interstellar beacons.

**Keywords:** Probes, SETI, ET, Altruism, Asteroids

## 1. INTRODUCTION

Since Frank Drake's first modern attempt to listen for artificial extra-terrestrial radio transmissions in 1960 [1], there have been many such searches [2, 3, 4]. Almost all have been attempts to detect electro-magnetic (EM) signals originating from interstellar space. Bracewell first proposed that ET might instead opt to send fully automated spacecraft, or probes, to our Solar System [5, 6], a theme further developed by Freitas [7, 8]. Rose and Wright [9] calculated the relative efficiency of probes when measured in cost per byte of information conveyed across the cosmos. Apart from some thinking on the subject in science fiction [10] not much more has been written on the subject, and, crucially, no serious, much less systematic search for probes has ever been conducted. However, were probes broadcasting to Earth from solar or terrestrial orbits, or from the surface of asteroids or moons, they might well have been detected serendipitously were they to have come within the same field of view of targeted SETI star searches, or they might have shown up as unexplained "noise" in various radio telescope surveys, Project SERENDIP [11], or the Deep Space Network [12].

Nonetheless, all prior SETI searches have been explicitly designed to either intercept inter- or intra-civilization communication streams, or to detect signals that are intentionally beamed at Earth, so called beacons. However, ET may actually prefer to send information physically via probes.

## 2. WHY SEND PROBES?

### 2.1 Reasons

- *Cost*: Probes are a much cheaper way to transmit cultural values and knowledge, provided that ET values time, labor, raw materials, and energy and the various overlaps among them in a manner similar to ourselves.
- *Time*: Probes are launched once (one and done); EM beacons, to be effective, must be transmitted continuously over millions or even billions of years.
- *Temporal efficiency*: If Earth is the only biologically indicative planet within ET's radius of interest, then their beacon might be pointed at Earth 24/7 over whatever time is required. However, if Earth is just one on a long list of targets of interest to ET, and ET's home star is just one on a large list of targets for our telescopes, unless ET builds at least one very powerful transmitter for each target, a temporal alignment between its transmitters and our receivers will become more unlikely as the number of targets per transmitter increases.
- *Labour:* Small probes might be built on an assembly line at less overall expense than the build out and maintenance of one or more gargantuan transmitters.
- *Raw materials*: With the same quantity of raw materials many probes might be built for each very large transmitter.
- *Energy:* The energy required to accelerate to only relatively low speeds (see below) probably from the surface of a low gravity object (e.g., space station, small moon or asteroid) of even very many probes would still be orders of magnitude less expensive than the energy needed to power transmissions over millions or billions of years.

### 2.2 Speed

There would be no need for speed in the event ET sent a probe as a response to its first detection of Earth's biologically indicative atmosphere. ET would correctly reason that Earth's bio-signature would precede the evolution of technologically competent intelligence by hundreds of millions or billions of years. Therefore, a probe could take an arbitrarily long time to arrive, and hence the energy requirements and cost of launch would be relatively small. In the grand scheme of things, it would not matter at all if a probe took even millions of years to arrive. A slower probe is actually preferred in terms of the energy required to decelerate it, as well as to reduce damage due to kinetic impacts with small dust particles en route.

* Foundation for Investing in Research on SETI Science and Technology (FIRSST), Berkeley, California.

### 2.3 Energetics

The exact comparison of economics and energetics between a probe and EM transmissions cannot now be calculated because it will depend on the as yet unknown mass of the probe, the speed to which it is accelerated, and the gravity at its site of launching, as compared to the signal strength, dwell time and duration of the beacon transmission. However, any reasonable estimate will greatly favor probes [13, 14, 9] such that the energy required to transmit ET's knowledge and culture by means of ET's equivalent of a hard drive delivered by probe should be vastly less than the energy required to transmit the same knowledge and culture using EM signals broadcast continuously over potentially billions of years.

Instead of broadcasting a beacon from ET's home planet using immense energy merely to attract our attention, a beacon from an orbiting probe (a) might be initiated only after it detects Earth's artificial EM leakage; (b) might require only a few watts of energy to broadcast its beacon; and (c) the intended recipients might receive the information bundle as soon as it has the technology to retrieve the probe, or, in the event the probe broadcasts its full message, to aim an appropriate receiver at it, rather than having to wait for potentially thousands of years for messages to travel back and forth across the Galaxy.

The current generation of radio telescopes can detect ET's interstellar radio beacons, but will not likely be able to extract a message that might be imbedded within it. To detect anything more than ET's carrier wave either we must build vastly larger radio telescopes, or ET must use a transmitter whose power vastly exceeds our current technology. Either way, a small probe is much more efficient, either because ET values its own expenditures or correctly anticipates the limitations of our receiving technology.

### 2.4 The L Factor

Probes solve for the problem of the *L* factor in the Drake equation, namely the length of time that a civilization broadcasts to the stars. If ET stops broadcasting in a short amount of time, either because it has lost interest, has self-destructed or has otherwise gone extinct, we would be very unlikely to receive its transmissions, simply because our civilization and theirs would not temporally co-exist. However, if ET communicates via probes, it need not have to live long enough to coincide temporally with mankind in order to transmit its desired information bundle to our younger civilization. By sending probes, ET might seek to insure that its knowledge and culture will survive its own demise.

### 2.5 It is Safer for ET to Send Probes

Probes avoid ET's reasonable fear that the recipient civilization (or an eavesdropper in the same communication pathway) may turn out to be hostile. If lacking such confidence, ET's probe need not reveal its position of origin. By definition, interstellar beacons do reveal their location. An intelligent probe might be capable of making its own determination as to whether the receiving civilization might be hostile, and be authorized to give out its return address only in the event that a threat is not perceived.

### 2.6 Probes as Life Forms

It is an oft-stated thesis of SETI scientists that ET is much more likely to broadcast than to visit. Interstellar space travel may be simply too onerous for muti-cellular carbon based life forms. For mankind, employing any technology we can currently envision, a trip to even the nearest stars would entail very many generations of whom all but the first and last would be born, live their lives and die en route. However, another oft-repeated thesis of SETI scientists is that the first life we encounter may actually be silicon rather than carbon based [15]. In effect, a biological life form would invent AI and upload its consciousness to it. That AI might self-replicate and very rapidly evolve in a non-Darwinian fashion. Such silicon beings would surely be better able to endure long space journeys than carbon based biology as we know it. Consequently, probes might not be simply ET's artifact, but its life form, fully able to pass the Turing test.

## 3. WHERE TO LOOK

The author believes that the most likely place in the Solar System to find probes is on the surface of asteroids. However, let us examine other possibilities first.

### 3.1 Earth

Earth would be a poor choice to look for a probe, since any that were sent here long ago would have by now been buried in sediment, subducted, inundated, covered by lava or glaciers, or stepped on by a dinosaur. Even whole cities that are a mere few thousands of years old are already buried deep in the dust. A probe sent to Earth millions or billions of years ago would probably never be found, but if they were to become the object of search, it should be by geologists or archeologists, and not by astronomers. Additionally, the machinery needed to affect a soft landing on a relatively large gravity, thick atmosphere object such as the Earth might add deleteriously to the size and complexity of the probe.

### 3.2 Lagrange Points

Lagrange points expose probes to sustained radiation, and would require a propulsion system built to last for eons in order to maintain orbital stability. A probe might require a propulsion system capable of decelerating and landing on a moon or asteroid, but that system need be engineered to be used just once.

### 3.3 Moons

A probe could bury itself on Earth's Moon to protect against radiation, but there are fewer useful raw materials, assuming that the probe is capable of self-repair or build out, and more energy would be required to decelerate on landing. Other moons in the Solar System may have a better assortment of useful materials near their surfaces, but many have active geologies (e.g., Io) or tectonic-like ice layers (e.g., Europa), that may bury or subduct probes over time, and relatively large gravities and atmospheres making landing more difficult and hazardous.

### 3.4 Oort Cloud and Kuiper Belt

Comets in the Oort cloud are too distant from the sun for the purposes of efficient PV energy generation. Also, ET's beacon would have to be much more powerful to be detected from Earth, and there would be trillions of targets for SETI scientists to sift through distributed in a full sphere. ET would hopefully have more sense than to sit on an Oort cloud object. Kuiper objects would suffer from similar, though lesser, limitations.

### 3.5 Asteroids

Inner Solar System asteroids are rich in raw materials and close enough to the sun for practical PV energy generation. They are typically much lower gravity objects than major moons, and devoid of atmospheres and active geology. Hence they are easier to land on, and probes would not be subjected to the subduction, inundation or other vicissitudes that would result from an active geology. Since they are largely restricted to the ecliptic they offer a smaller search area than the Oort cloud. There are also far fewer asteroids than Oort or Kuiper objects, making such a search easier. A probe could burrow into an asteroid for long term refuge from radiation, emerging very briefly, for example, only once a century, to monitor for artificial EM leakage, which would, if detected, trigger its beaconing behavior. Therefore, the author believes that asteroids are the best candidates for probe searches.

## 4. WHAT TO LOOK FOR

As pure speculation, the author would look for a low power (e.g., <10W) radio beacon. At some frequency or frequencies, the probe might, in effect, simply "beep! beep! beep!". In order to download the probe's useful information it would have to be physically retrieved. In theory, ET probes could transmit using optical lasers. However, with so many amateur and professional telescopes trained on the sky every night, it would seem that ET lasers should have serendipitously shown up on some CCD outputs by now.

The probe itself would likely be small but much larger than its payload alone, that payload being its information storage device, equivalent to a hard drive. However, the probe would also require shielding, a guidance system, sensors, deceleration and landing capability, transmitter, and fabrication capability akin to a 3-D printer to repair itself and/or build out new capabilities or even to replicate in situ. As pure speculation, this author imagines that probes might be larger than a basketball, but smaller than an automobile.

## 5. PROBES MAY BE PLENTIFUL

If extant, probes may be far more plentiful than interstellar beacons. There are three possible pathways by which the Solar System might be rich in probes:

### 5.1 Multiple Civilizations

Probes may have been sent by multiple civilizations, many or most of whom could well be currently extinct. Unless $L$ is quite large, this by itself might mean that probes would be far more numerous than interstellar beacons.

### 5.2 Sequential Versions

Some or all ET civilizations might have sent multiple probes sequentially, as versions 1.0, 2.0, 3.0, etc., each informing us of the latest developments in their arts, history, philosophy, science, or whatever they wish to share. If we ourselves had sent probes, say, 200 years ago, informing our galactic neighbors of the great achievements of Newton, Rembrandt and Beethoven, then a mere 200 years later we would probably feel impelled to send out new probes, now featuring Einstein, Tarantino and Elvis.

### 5.3 Replication upon Arrival

Some or all of the probes might be von Neumann replicators [16], that is, they may duplicate themselves upon arrival using the raw materials they find here.

## 6. IF PLENTIFUL, WOULD PROBES BE EASY TO FIND?

A crucial feature of probes used as beacons is that they are, by definition, and unlike their interstellar counterparts, always pointed at Earth. Logically, they should have a target set of one. By contrast, interstellar beacons may cycle sequentially through a list encompassing hundreds, thousands or millions of targets.

Over possibly a billion years or more, probes might really pile up, such that there may be very many of them in the ecliptic, possibly designed to be easily detectable. Testing the probe hypothesis might not require our most powerful telescopes. Even small radio telescopes might be able to easily detect broadcasting probes.

However, this line of reasoning can be completely turned around. Given that no probe has heretofore been discovered by serendipity, this might only mean that (a) probes are not plentiful; or (b) most probes do not continuously transmit to Earth. Maybe they are plentiful, but do not broadcast unless first awakened by our own transmissions, perhaps by radar. Priority might therefore be given to turning radio telescopes toward asteroids that have previously been probed by radar. Perhaps probes transmit very intermittently, for example, only once per century. Probes might be plentiful but most or all do not broadcast. Rather, they might have to be physically encountered. In such event, their discovery will have to await a time when spacecraft routinely visit asteroids, presumably on mining missions. More cynically, as Brin has argued [17], perhaps probes are "lurkers," intentionally remaining silent for purposes of their own, such as to "steal" our science and culture while giving nothing in return. Such spy probes might be intentionally designed to avoid detection, making their discovery by any current technology very unlikely. Ominously, it has also been suggested that ET probes may be in our Solar System now, however, neither for the purpose of communicating with us nor to spy on us. Rather, their mission is to destroy Earth at their first detection of technologically generated EM. They might, therefore, already be at their nefarious work of redirecting the orbits of comets or asteroids to impact our planet [18].

## 7. WHAT'S IN IT FOR ET?

The main drawback that ET might consider in sending probes is that it will assume, *a priori*, that either there will never be a response, or a response will be delayed by many millions of years. What's in it for ET under these circumstances? The probable galactic currency is knowledge. Using EM transmissions, there can be commerce, albeit, delayed by the speed of light. With commerce as a motivation, trade might look something like this: ET might transmit an abridged version of its knowledge and culture as a loss leader. When Earth transmits in return, say, some Mozart, Maxwell's equations, Spielberg's *Close Encounters of the Third Kind,* and information on a few hundred species, ET will send more information. If we next send Beethoven, Bach, the Beatles, Einstein's equations, Monet, and the human genome, ET will send more. Finally, once trust has been established, and we give them everything (the Internet, Library of Congress, all of our music, etc.), then ET will send its unabridged version of Encyclopedia Galactica (EG). Using radio or optical transmissions, even if an ET at 500 LY distance must wait 1000 years for a response, at least there will be one. A fairly complete exchange, involving practices of fair trade, might take less than 10,000 years. So why would ET choose to send information laden probes if they understand, *a priori,* that the best they can hope for is a first response hundreds of millions or billions of years into their future, when allowing for the lag time between their first detection of our biological atmosphere and the development of radio and optical transmitters by that biology? Possibly, there will not be much beyond the altruistic joy of spreading its culture and knowledge. Or perhaps there would be a larger payoff. For example, ET could transmit its genomic sequence with a request that we resurrect it in our Solar System. In fact, they could demand this in exchange for the password to their probe's information payload.

Would ET send EG altruistically? Perhaps. Frank Drake and Carl Sagan spent time, energy, raw materials and labor constructing their famous Voyager plaques for a slow one way mission to nowhere in particular. Missionaries have gone to the far reaches of the Earth, and in great discomfort, to transmit their culture and religion. Moreover, an ET civilization need not be infinitely altruistic to send probes. It would take just one of their "billionaires" or "potentates" in just one of their generations to do it just once. Unlike radio or optical transmissions, probes do not require a sustained millions or billions of year's dedication to the enterprise. Finally, *L* may be the great motivator. Any civilization facing its own mortality might send probes as a way to perpetuate its achievements. Would we not also send such a message in a bottle were we facing our certain doom? Would we allow all that we have achieved to just vanish?

## 8. CONCLUSIONS

The most optimistic case for probes is unlikely to be true, namely, that vast numbers of probes are in the asteroid belt constantly beaming beacons towards Earth in order to attract our attention. In such an event, by serendipity, a detection would likely have occurred by now. Nevertheless, since targeted SETI searches have almost always been aimed at stars, with a foreground intra-Solar System probe detection possible only by serendipity, the author recommends that some search time be allotted to asteroids. Such a search would, per force, still target stars, namely those that happened to be in the background, and far more still if the selected asteroids were located within the plane of the Milky Way. However, in that case it would be the interstellar ET signal that would be detected by serendipity.

Regarding asteroids as potential targets will require some additional shifts in search strategies. Signal detection algorithms would need to account for a drift across the field of view beyond the sidereal stellar rate. If ET's beam is narrow and not fixed on a single Earth coordinate, it will appear to blink off as the asteroid rotates. It will also blink off in any SETI search that slews siderally with the fixed stars as the asteroid passes through the telescope's field of view. The famous 1977 WOW signal blinked off after 72 seconds and was not observed again in the same location. Crucially, when astronomers record a transient signal in their data that cannot be dismissed as RFI, they should not only return for follow up observations to the static coordinates, as though the signal derived from a star, but also determine what asteroid(s) was in the field of view at the time of the detection, and observe at the coordinates of the asteroid's then current location. Such follow up observation should dwell on the asteroid for at least one of its rotations, if the rate is known, or, otherwise, at least for the duration of the average rotation rate of similar objects.

* * *